\documentclass[copyright,creativecommons]{eptcs}
\providecommand{\event}{RTRTS 2010}
\usepackage[english]{babel}
\usepackage[latin1]{inputenc}
\usepackage{amssymb}
\usepackage{lmodern}
\usepackage[T1]{fontenc}
\usepackage{amsmath}
\usepackage{bussproofs}
\usepackage{tikz}
\usepackage{cite,url,xspace,courier} 
\usepackage{creol}
\usepackage{comment}
\usepackage{multirow}

\usepackage{breakurl}

\newif\ifextracted
\extractedtrue

\newcommand{\TYPE}[1]{\mbox{\sffamily #1}}
\newcommand{\key}[1]{\mbox{\ttfamily\bfseries #1}}
\newcommand{\id}[1]{\mbox{\itshape #1}}
\newcommand{\many}[1]{\{#1\}}
\newcommand{\option}[1]{[#1]}

\newcommand{\Creol}{\mbox{Creol}\xspace}

\lstdefinelanguage{Maude}{
  keywords={sort,sorts,subsort,subsorts,var,vars,assoc,associative,ceq,cq,%
            comm,commutative,config,configuration,ctor,constructor,ex,%
            extending,id:,identity,identity:,idem,idempotent,inc,including,%
            iter,iterated,owise,otherwise,poly,polymorphic,prec,precedence,%
            pr,protecting,strat,strategy,op,ops,eq,label,to,is,mod,endm,fmod,%
            endfm,rl,crl,view,endv,eof,load,in,include,if,then,else,fi,from,by,%
            and,not},
  sensitive=true,
  morecomment=[l]{--},
  morecomment=[l]{***},
  morecomment=[s]{***(}{)},
  morestring=[b]",
  literate={+}{{$+$}}1 {/}{{$/$}}1 {*}{{$*$}}1 {=}{{${}={}$}}1
           {>}{{$\!\!\rangle\;$}}3 {<}{{$\langle\,$}}1 {|}{{$\!\!\mid\:$}}1 {:}{{$\colon$}}1
           {->}{{$\rightarrow\;$}}2 {>=}{{$\geq$}}2 {<=}{{$\leq$}}2
           {=>}{{$\longrightarrow\;$}}2 {--}{-{}-}2 {:=}{{$:=$}}2
           {~>}{{$\leadsto$}}2 {==}{{${}={}$}}2
           {|->}{{$\mapsto$}}3 {---}{-{}-{}-}3 {=/=}{{${}\ne{}$}}1
}

\def\codesize{\fontsize{8}{9}}

\lstnewenvironment{maude}{%
    \lstset{language=Maude,columns=fullflexible,mathescape=true,%
            keywordstyle=\bfseries,commentstyle=\sl\ttfamily,%
            basicstyle=\codesize\tt,inputencoding=latin1, extendedchars}%
   \csname lst@SetFirstLabel\endcsname}
  {\csname lst@SaveFirstLabel\endcsname}

  \title{Lightweight Time Modeling in Timed Creol\thanks{This research
      was done in the context of the EU project FP7-231620 \emph{HATS}: Highly
      Adaptable and Trustworthy Software using Formal Methods
      (\url{http://www.hats-project.eu}).}}%
  \author{Joakim Bj{\o}rk \quad Einar Broch Johnsen \quad Olaf Owe \quad
    Rudolf Schlatte \institute{Department of Informatics, University of
      Oslo, Norway} \email{\{joakimbj,einarj,olaf,rudi\}@ifi.uio.no} }

\def\titlerunning{Lightweight Time Modeling in Timed Creol}
\def\authorrunning{J.\ Bj{\o}rk, E.~B.\ Johnsen, O.\ Owe \& R.\ Schlatte}
\begin{document}

\lstset{language=Creol,columns=fullflexible,basicstyle=\ttfamily,mathescape}

\maketitle

\begin{abstract}
  \Creol is an object-oriented modeling language in which inherently
  concurrent objects exchange asynchronous method calls.  The
  operational semantics of \Creol is written in an actor-based style,
  formulated in rewriting logic. The operational semantics yields a
  language interpreter in the Maude system, which can be used to
  analyze models.  Recently, \Creol has been applied to the modeling
  of systems with radio communication, such as sensor systems.  With
  radio communication, messages expire and, if sent simultaneously,
  they may collide in the air.  In order to capture these and other
  properties of distributed systems, we extended \Creol's operational
  semantics with a notion of time.  We exploit the framework of a
  language interpreter to use a lightweight notion of time, in
  contrast to that needed for a general purpose specification
  language.  This paper presents a timed extension of \Creol,
  including the semantics and the implementation strategy, and
  discusses its properties using an extended example. The approach can
  be generalized to other concurrent object or actor-based systems.
\end{abstract}

\section{Introduction}
\label{sec:intro}
Actor-based systems consist of autonomous ``actors'' with explicit
identity, which execute local tasks and asynchronously exchange
messages \cite{Agha86,AMST97}. Actor-based systems are attractive for
the modeling of distributed computing systems due to their separation
of concerns between local computation on the one hand and
communication and synchronization on the other hand, which fits with a
natural way of understanding distributed systems. \Creol is an
object-oriented modeling language in which inherently concurrent
objects exchange asynchronous method calls
\cite{johnsen07sosym}. The semantics of \Creol is defined
in rewriting logic \cite{Meseguer92}; in fact, we have used Maude
\cite{Clavel02} as an underlying simulation platform for \Creol models
and extended the interpreter for visualizing the runtime state as well
as for dynamic symbolic execution \cite{griesmayer09tap}.
A novel feature of \Creol is that method activations may explicitly
suspend execution while waiting for some condition; e.g., the answer
to a method call. This results in a very intuitive model of a
computation unit which combines active and reactive behavior.
In the semantics of \Creol, asynchronous method calls are encoded
using asynchronous message passing. Ignoring object-oriented features
such as the late binding of method calls, a concurrent object in
\Creol may be understood as an actor in which tasks are executed
using cooperative scheduling.
Due to its close relationship with actor systems, \Creol is a
concurrent imperative language which actually supports compositional
reasoning \cite{deboer07esop}, in contrast to object-oriented
languages like, e.g., Java.

In many cases, time influences the desired (or actual) behavior of
systems. As an interesting example, wireless sensor systems behave
according to timing properties. The radio unit of a sensor typically
has different modes; the radio may be receiving, sending, or
dormant. Unless the radio is in receiving mode, a message sent to the
sensor will be lost. If two concurrently running sensors send messages
at the same time, the messages may interfere with each other and
their content is lost. These properties of wireless sensor networks
introduce many interesting challenges for formal methods. A sensor
radio with explicit timing parameter settings has been modeled and
analyzed in Uppaal \cite{tschirner08bsn}. A plethora of new network
protocols have been proposed for wireless sensor networks, due to
their ad-hoc and self-organizing nature. \"Olveczky et al.\ have shown
how such protocols may be modeled and analyzed
\cite{olveczky09journal,katelman08fmoods}, based on Real-Time Maude
\cite{olveczky02tcs,olveczky07hosc}.

For \emph{timed distributed systems}, time is either modeled by a
global clock (or equivalently, local clocks which evolve with the same
rate), or by local clocks.  For simplicity we use a so-called
\emph{fictitious clock model}~\cite{alur92:_logic_model_real_time}
based on a global clock, which allows us to ignore clock
synchronization between objects.  When modeling timed systems, one may
consider different time domains from a partial ordering of events, via
a discrete time domain to the detailed continuous time. Increasing the
level of granularity of the time domain leads to an increased
complexity of the models.  For our purpose of grouping simultaneous
(communication) events with an interleaving semantics, it suffices to consider a
discrete time domain.
The values need not correspond to values of real clocks, and are
therefore usually chosen to be natural numbers that count steps.
Effects such as radio broadcast are confined to a particular instance
of time, and broadcast messages disappear as soon as time advances.  


In this paper we present a timed version of Creol based on a
fictitious clock model. The model is without local clocks.  At the
language level no additional syntax is needed apart from read-only
access to the global clock, using the variable \emph{now}. The major
argument for keeping the time model lightweight is to keep the state
space as small as possible for model checking purposes.  As a case
study we present a model of a wireless sensor network together with
execution results.
We will compare our time model
for Creol to other formalisms such as Real-Time Maude.

The paper is structured as follows: Section~\ref{sec:creol:intro}
presents the modeling language \Creol. In Section~\ref{sec:creol:time}
\Creol is extended with time. Both syntax and semantics are
presented. An example is shown in
Section~\ref{sec:example}. Section~\ref{sec:related-work} gives a
comparison of timed \Creol to other timed
models. Section~\ref{sec:conclusions} concludes the paper.

\section{A Short Introduction to \Creol}
\label{sec:creol:intro}

We first introduce the features of the object-oriented modeling
language \Creol which are necessary to understand the approach
presented in this paper.  A more detailed introduction to \Creol can
be found in, e.g.,~\cite{johnsen07sosym,deboer07esop}.

\Creol\ features imperative programming constructs for distributed
active objects, based on asynchronous method calls and processor
release points. Asynchronous method calls may be seen as triggers of
concurrent activity, resulting in new activities (processes) within
the called object.  Objects are dynamically created instances of
classes, an \emph{init} method is used to initialize the object's
fields at creation time. Active objects encapsulate a current activity
(a thread) and an internal activity pool.  Active behavior, triggered
by a \emph{run} method, is interleaved with passive behavior
(triggered by method calls) by means of processor release points. At
each point in time at most one thread is active in each object. The
scheduling of threads is by default non-deterministic, but more refined
scheduling strategies may be given, as in~\cite{deboer09fsen}. The
modeling language includes a functional expression language for values
of basic data types, which will not be explained in detail.  Objects
are uniquely identified; communication takes place between named
objects, and object references may be exchanged between objects.
Object variables are typed by interfaces.  The language is strongly
typed: for well-typed programs, invoked methods are supported by the
called object (when not \id{null}), such that formal and actual
parameters match.  This also includes call-backs from objects to their
environment via the special \emph{caller} variable.

\begin{figure}[t]
  \small
  \begin{equation*}
    \begin{array}{cl}
      \begin{array}{l}
        \emph{Syntactic}\\
        \emph{categories.}\\
        C, I, m \textrm{ in } \TYPE{Names}\\       
        t \textrm{ in } \TYPE{Tag}\\
        g \textrm{ in } \TYPE{Guard}\\
        s \textrm{ in } \TYPE{Stmt}\\
        x \textrm{ in } \TYPE{Var}\\
        e \textrm{ in } \TYPE{Expr}\\
        o \textrm{ in } \TYPE{ObjExpr}\\
        b \textrm{ in } \TYPE{BoolExpr}
      \end{array}
      \quad
      &
      \begin{array}{rrl}
        \multicolumn{3}{l}{\emph{Definitions.}}\\
        \mathit{IF}&::=&
        \key{interface}\ I\ \key{begin}\ 
        \option{\key{with}\ I\ \many{Sg}}\ \key{end}\\
        CL &::=& \key{class}\,C\,
        \option{(\many{x :I})}\\
         &&\option{\key{implements}\,\many{I}}\,\key{begin}\
        \many{\key{var}\,x:\id{I}\,\option{:=e}}\,
        \many{\option{\key{with}\,I}\,\many{M}}\,\key{end}\\
        \textsl{M} &::=& \textsl{Sg} == \option{\key{var}\ \many{\many{x} :\id{I}\, \option{:=e}};}\ s\\
        \textsl{Sg} &::=& \key{op}\ m\ (\option{\key{in}\ \many{x :I}} \option{\key{out}\ \many{x :I}})\\
        g&::=& b \mid t? \mid g\land g \mid g\lor g\\
        s&::=& s;s \mid s [] s \mid x:= e
        \mid x:= \key{new} \ \id{C}\ \option{(\many{e})}\\
        &|& \key{if}\ b\ \key{then}\ s\ \option{\key{else}\ s}\ \key{end}
        \mid \key{while}\ b\ \key{do}\ s\ \key{end}
        \mid \key{await}\ g \\
        &|& \option{t}!\option{o.}m(\many{e})  \mid t?\option{(\many{x})}\mid
        \option{\key{await}}\option{o.}m(\many{e};\many{x})         
        \strut
      \end{array}
    \end{array}
  \end{equation*}
  \caption{A simplified language syntax. Terms such as $\many{e}$ and
    $\many{x}$ denote lists over the corresponding syntactic
    categories and square 
    brackets denote optional elements.}
  \label{fig:creol-comps:bnf}
\end{figure}

\paragraph{Basic statements.}

Figure~\ref{fig:creol-comps:bnf} displays a simplified formal syntax of
\Creol\ programs. Inheritance in both interfaces and classes is omitted
for brevity. A program consists of interface and class definitions.
Classes $CL$ contain definitions of attributes $x$ (with initial values)
and methods $M$.  A method contains a list of local variable
declarations, and a statement $s$, which may
access class attributes, locally defined variables, and the formal
parameters of the method (given after the keywords \key{in} and
\key{out}).  An interface definition $\mathit{IF}$ contains method
signatures $\mathit{Sg}$ associated with \emph{co-interfaces} $I$ given
by a \key{with} clause.  The co-interface $I$ specifies the type of a
client of $\mathit{IF}$ and enables to express requirements on
call-backs.  Finally, a class implements a list of interfaces, thereby
specifying the type of its instances. In order to allow call-backs, a
method may use the implicit \emph{caller} parameter typed by the
co-interface of the method.  Class parameters, input parameters, and the
self-reference \emph{this}, are read-only.  Remote access to attributes
is not allowed, method interaction is the only means of communication
between objects.  Assignment, \key{if}-, and \key{while}-constructs are
standard. The box \texttt{[]} is the non-deterministic choice
operator. Creol also includes standard string, numeric, and Boolean
datatypes, as well as lists, sets, maps and tuples, and their standard
operators.

The guard $g$ controls processor release in the statement
$\key{await}\ g$, and consists of Boolean conditions that contain
return tests (see below).  If $g$ evaluates to false, the current
activity is \emph{suspended} and the execution thread becomes
idle. When the execution thread is idle, any enabled activity may be
chosen from the pool of suspended activities.  Explicit signaling is
therefore redundant.  The \emph{run} method of an object is called
after initialization, and initiates active behavior.  Release points
in the run method allow activities in the activity pool to be
scheduled.

\paragraph{Communication.}

After making an asynchronous method call $t!o.m(\many{e})$, the caller
may proceed with its execution without blocking on the method reply.
Here ${o}$ is an object expression and $\many{e}$ are (data value or
object) expressions.  The tag $t$ will be assigned a unique value that
identifies the call, which may later be used to refer to that call in
two different ways: First, the guard $\key{await}\ t?$ suspends the
active activity unless a return to the call associated with $t$ has
arrived.  Second, the return values are retrieved by the 
\emph{reply statement} $t?(x)$, which is blocking the object until the
return values are present.  Local calls are written $t!m(\many{e})$.
If no return values are desired by the caller, the tag may be omitted;
e.g., $!o.m(\many{e})$.  The sequence $t!o.m(\many{e});\ t?(x)$
encodes a \emph{blocking call}, abbreviated $o.m(\many{e};x)$ (often
referred to as a synchronous call), whereas the call sequence
$t!o.m(\many{e});\ \key{await}\ t?;\ t?(x)$ encodes a non-blocking,
\emph{preemptable call}, abbreviated $\key{await}\ o.m(\many{e};x)$.

\section{A Time Model for \Creol}
\label{sec:creol:time}

In order to reason about time in a Creol model, the concept of time
has to be introduced to the language.  Our timed Creol interpreter
adds a datatype \texttt{Time} and its accompanying operations to the
language.

A value of type \texttt{Time} can be obtained by evaluating the
expression \key{now}, which returns the current time, i.e.,\ the value
of the global clock.  Given a time value, other time values can be
constructed by adding and subtracting duration values.

Time values form a total order, with the usual less-than operator
semantics.  Hence, two time values can be compared with each other,
resulting in a Boolean value suitable for guards in \key{await}
statements.  While all other time values are constant, the result of
comparing the expression \key{now} with another time value will
change with the passage of time.  As an example, given a tag
\texttt{l}, the following Creol fragment
\begin{creol}
  var t : Time := now;
  await l? && now < t + 10 ; SL
  []
  await now > t + 10 ; EL
\end{creol}
models both the normal (\texttt{SL}) and the timeout (\texttt{EL})
behavior of the synchronization with the method invocation associated with
\texttt{l}.  The box \texttt{[]} is the non-deterministic choice
operator, which chooses one of its two statements if both are enabled,
and blocks until at least one statement is enabled.  (In the example
above, the choice operator does not add non-deterministic behavior to the
model, since the two guards are mutually exclusive.)

In this model of timed behavior, the passage of time needs never be made
explicit in the model, as with e.g.\ a \texttt{tick} statement.
Instead, passage of time is observed within \key{await} statements,
and time is advanced when no other activity may occur.  Note, though, that
the semantics of this model of time, combined with \Creol's blocking and
non-blocking synchronization semantics, are powerful enough to express
both activity- and object-wide \texttt{tick} statements: given the
following method definition:
\begin{creol}
op tick(in duration: Int) == var t : Time := now ; await now >=  t + duration$,$
\end{creol}
The following two lines of timed \Creol express activity- and
object-wide \texttt{tick}, respectively, by using non-blocking and blocking synchronization on the tag \texttt{l}:
\begin{creol}
l!tick(1); await l?
l!tick(1); l?
\end{creol}

The remainder of this section describes the implementation of this Creol
time model as an operational semantics expressed in Maude.

\subsection{Operational semantics}
\label{sec:oper-semant}

We introduce a clock that holds the current global time value.  This
value is accessible to Creol models via the \texttt{now} expression.
The global clock is a structure containing an identifier and two
natural numbers:
\begin{maude}
< O : Clock | time: T, limit: B >
\end{maude}
The first number, \texttt{T}, is the current time.  
\texttt{B} provides an upper time bound for model
execution. \texttt{O} is a unique identifier for this clock
object. The identifier is superfluous as models should only contain one
clock object, but is included to allow the clock to follow standard
object notation in Maude.

Each Creol object is represented as a Maude structure containing
instance variable bindings, the currently running process, the
process queue, and some housekeeping values.  Maude equations and rewrite rules
operate on these structures and implement the operational semantics of
Creol.

As an example, here is an outline of the rule in the untimed interpreter
that implements the caller side of asynchronous method calls (simplified
for demonstration purposes -- we leave out some details, such as
generating the unique identifier \texttt{N} for the fresh future
structure):
\begin{maude}
rl
  < O : C | Att: A, Pr: { L | F := call(O1, M, PL) ; SL }, PrQ: W >
=>
  < O : C | Att: A, Pr: { L :: F |-> N | SL }, PrQ: W >
  < N : Future | Completed: false, Ref: 1, Value: emp >
  invoc (O1, M, eval(PL, A :: L), O, N)
[label async-call] .
\end{maude}
The object \texttt{O} of class \texttt{C} (with attributes \texttt{A})
has an active process with a state \texttt{L} and a first statement
\texttt{call} to the method \texttt{M} of object \texttt{O1}.  The
result of that statement is to be stored in the local variable
\texttt{F}.  The rule \texttt{async-call} removes the call statement
from the active process and updates the binding of the tag \texttt{F}
with a reference to a new future \texttt{N} that holds the status of the
method invocation (completed or not) and its return value.  (A separate
structure is necessary since Creol's Futures are first-class values that
can be referenced from multiple objects.)  Finally, the rule also
generates an \texttt{invoc} structure for \texttt{O1} that will, in the
rule \texttt{message-receive}, cause a process to be inserted into the
callee's (\texttt{O1}'s) process queue \texttt{PrQ}.

\begin{maude}
rl
  < O : C | Att: A, Pr: P, PrQ: W >
  < C : Class | Mtds: (MS, < M : Method | Param: L, Code: SL ) >
  invoc (O, M, DL, O1, N)
=>
  < O : C | Att: A, Pr: P, PrQ: (W, { L $\mapsto$ DL, caller $\mapsto$ O1, .result $\mapsto$ N | SL } ) >
  < C : Class | Mtds: (MS, < M : Method | Param: L, Code: SL ) >
[label message-receive] .
\end{maude}

The \texttt{message-receive} rule, presented above, appends a new
process into the receiving object's process queue.  Method lookup is
done by name in the class of the object (we elide handling of method
inheritance for reasons of brevity); argument values \texttt{DL} are
bound to the method's parameters \texttt{L}.  Additionally, the
\texttt{caller} attribute is set to a reference of the calling object
\texttt{O1} and an internal field \texttt{.result} stores a reference to
the future \texttt{N} that will store the result of the method
invocation.  Finally, the statement list of the process is initialized
with the statement list \texttt{SL} of the method definition.

All rules for executing statements follow this general pattern.
Specifically, note that these rules can execute mostly independently of
the global clock value.  The only exception is the \Creol expression
\lstinline[language=Creol]!now!  which, given the presence of a clock
\lstinline[language=Maude] !< O : Clock | time: T, limit: B >!, will
evaluate to a term \lstinline[language=Maude]!time(T)!.  This in turn
will influence Boolean guards involving the global time, hence the
readiness of processes waiting on such a guard becoming true.

For instance, the timed version of the rule async-call looks as follows:
\begin{maude}
rl
  < O' : Clock | time: T, limit: B >
  < O : C | Att: A, Pr: { L | F := call(O1, M, PL) ; SL }, PrQ: W >
=>
  < O' : Clock | time: T, limit: B >
  < O : C | Att: A, Pr: { L :: F |-> N | SL }, PrQ: W >
  < N : Future | Completed: false, Ref: 1, Value: emp >
  invoc (O1, M, eval(PL, A :: L, T), O, N)
[label timed-async-call] .
\end{maude}
The only difference to the untimed version is that the \texttt{eval}
function gets an additional argument, namely the time value.  Rule
\texttt{message-receive} does not need to be changed for timed \Creol at
all, since no evaluation of expressions takes place in that rule.

Consequently, for the timed interpreter the global clock has to be added
both to the left-hand and right-hand side of any rule that involves
evaluating an expression.  Here is the rule for evaluating an
\lstinline[language=Creol]!await! statement in the timed interpreter:

\begin{maude}
crl
  < O : C | Att: A, Pr: { L | await E ; SL }, PrQ: W >
  < O' : Clock | time: T, limit: B >
  CN
=>
  < O : C | Att: A, Pr: { L | SL }, PrQ: W >
  < O' : Clock | time: T, limit: B >
  CN
  if evalGuard(E, (A :: L), CN, T) asBool
[label await] .
\end{maude}
If the condition \texttt{E} holds, as determined by the auxiliary
equation \texttt{evalGuard}, then the
\lstinline[language=Creol]!await!  statement simply reduces to a
\texttt{skip}. In contrast to the \texttt{eval} function above,
\texttt{evalGuard} may need to consider futures in the environment,
and gets the additional parameter \texttt{CN}.  In contrast to the
\texttt{tick} rule presented below, there is no necessity of
referencing the whole system in \texttt{CN} here, since Maude will
choose a subset as needed.  Another rule in the interpreter, not
presented here, has the task of suspending the active process if
\texttt{E} evaluates to \texttt{false}.

\paragraph{Clock advancement.}
The global clock cannot advance freely.  We specify run-to-completion
semantics, requiring all objects to finish their actions as soon as
possible.  Hence, there are several restrictions to when the global
clock may advance.  Clock advancement is blocked if:
\begin{itemize}
\item An object has an active non-blocked process.
\item Some process in the process queue of an object is enabled and
  ready to run.
\item A message (method invocation) is ``in flight'' between objects.
\end{itemize}

We define a Maude operator \emph{canAdvance} that traverses the
configuration of objects and returns \emph{false} if any of these
conditions are true.  A sketch of the its semantics, with some
implementation details elided, looks as follows:
\begin{maude}
op canAdvance : Configuration  Time-> Bool .

eq canAdvance (< O : C | Att: A, Pr: P, PrQ: W > CN, T) = 
  blocked(P, A, CN, T) and canAdvance(CN, T) .

eq canAdvance (< O : C | Att: A, Pr: idle, PrQ: W > CN, T) =
  allBlocked(W, A, CN, T) and canAdvance(CN, T) .

eq canAdvance (< M:Msg | > CN, T) = false .

eq canAdvance (CN, T) = true [owise] .
\end{maude}
The \texttt{blocked} operator reduces to \emph{true} if the process
\texttt{P} given as its first argument is not enabled.
\texttt{allBlocked} returns \emph{true} if all of the processes in the
process queue \texttt{W} are blocked.  Both these operators need the
current configuration to determine the status of processes currently
waiting on futures, which is why the configuration is an argument to
\texttt{blocked} and \texttt{allBlocked}.

With the \texttt{canAdvance} equation, the tick rule of the global clock
is straightforward. We increase the time by one time unit if the
clock can advance, as long as the time limit is not exceeded. The
definition of the tick rule is shown below (with \texttt{\{\}} enclosing
the whole configuration to produce a system, in the same manner as in
Real-time Maude).

\begin{maude}
crl
  { CN < O : Clock | time: T, limit: B > }
=>
  { CN < O : Clock | time: T + 1, limit: B > }
if canAdvance (CN, T) and T $<$ B
  [label tick] .
\end{maude}
Here the importance of the \texttt{limit} attribute of the clock becomes
clear: If all object activity has ceased, the clock is free to advance
without any bound.  The limit, which can be arbitrarily large, ensures
that an unbounded Maude rewrite command terminates.

\section{Example: A Model of a Wireless Sensor Network}
\label{sec:example}

We used timed Creol to model the behavior of a wireless sensor
network.  A typical sensor network consists of a number of sensors,
and a sink which collects data.  The sensors record some sort of data
and send it towards the sink via wireless links.  Due to the limited
power of the wireless signals, some sensors may not be directly
connected to the sink; these sensors are dependent on other sensors to
forward their messages.  The routing of messages towards the sink may
be done in many different ways to limit time consumption, energy
consumption, number of messages sent etc.  The example in this section
uses a simple flooding algorithm: when a sensor senses data, it
broadcasts it to all other nodes within range.  When a sensor receives
a message that it has not seen before, it is rebroadcast to all its
neighbors.  More involved routing algorithms exist where sensors
selectively rebroadcast messages depending on whether they are on the
path to the sink, but this simple algorithm suffices to show our
results. A more complex routing protocol modeled in \Creol is
investigated in~\cite{Leister1022}.

In our model, the nodes are not directly connected to each other.
Instead, each node has a reference to a \emph{Network} object which
models the behavior of the transmission medium between nodes.  This
structure enables us to model collisions, message loss, selective
retransmission, and the node topology (which nodes can be reached from
each node) without local knowledge inside the node objects.

\begin{figure}[t]
\begin{creol}
interface Node
begin
  with Network
    $\textsl{// Receive data from network.  data format: (id of originator, sequence no)}$
    op receive(in data: [Int,Int])
end

interface Network
begin
  with Any
    $\textsl{// Register `node' as part of the network and as able to send to the}$
    $\textsl{// nodes in `connections'}$
    op register(in node: Node, connections: List[Node])
  with Node
    op broadcast(in data: [Int,Int])
end
\end{creol}
  \caption{The interfaces of the sensor network}
  \label{fig:interfaces}
\end{figure}

Figure~\ref{fig:interfaces} shows the interfaces of both the nodes and
the network.  The \texttt{broadcast} method of the network gets called
by nodes when they wish to broadcast data; the \texttt{receive} method
of a node is called by the network with data that is broadcast by
another node.

The \emph{run} method of a \emph{Main} class (not given here) sets up
the sensor network by creating all \texttt{Sensor} objects, one
\texttt{Network} object, and establishing the topology.  The sensors are
created by statements such as \lstinline[language=Creol]!n1 := new SensorNode(1, nw, 3);!.
This statement results in the creation of a \emph{SensorNode} object
with \emph{id} 1. It will belong to the network \emph{nw} and do three
sensings. Later the sensor will be registered in the network by a
command like \lstinline[language=Creol]!nw.register(n1, [n2,n3];);!
which states that the sensor \emph{n1} should have a link to the nodes
\emph{n2} and \emph{n3}. These links are not necessarily symmetrical; it
is possible for a sensor to receive messages from another sensor but not
be able to send to it in turn.

\begin{figure}[t!]
\begin{creol}
class SensorNode(id: Int, network: Network, noSensings: Int)
implements Node
begin
  var received: List[[Int,Int]] := nil;  $\textsl{// All previously received messages}$
  var sendqueue: List[[Int,Int]] := nil; $\textsl{// Messages waiting to be sent}$
  var seqNo: Int := 0;                 $\textsl{// Running package seq. no}$
  var sending: Bool := false;
  
  $\textsl{// Forward (or send) a single message from the queue}$
  op sendOrForward == $\label{line:node:send}$
    var t: Time := now;
    var l: Tag[ ];
    await sending = false;
    sending := true;
    l!network.broadcast(head(sendqueue));
    sendqueue := tail(sendqueue);
    await l?;
    await now > t; $\label{line:node:timing}$
    sending := false

  op store(in data: [Int,Int]) ==
    sendqueue := sendqueue |- data

  $\textsl{// Produce a sensing (of the environment) and store it locally}$
  op sense == $\label{line:node:sense}$
    store((id,seqNo););
    seqNo := seqNo + 1

  op run == $\label{line:node:run}$
    await start = true;
    while true do
      await seqNo $<$ noSensings; sense(;)
      []
      await #(sendqueue) > 0; sendOrForward(;)
    end
  with Any
    op start == start := true

  with Network
    op receive(in data: [Int,Int]) ==
      await start = true;
      if ~(data in received) then $\textsl{// re-send if not seen before}$
        received := received |- data;
        store(data;)
      end

end
\end{creol}
 \caption{The implementation of the sensor nodes}
 \label{fig:sensor}
\end{figure}

When an object of class \emph{SensorNode}, shown in
Figure~\ref{fig:sensor}, is generated, its \texttt{run} method executes
(shown in line~\ref{line:node:run} ff.), choosing one of two behaviors.  If
the sensor has not yet done the number of sensings it is supposed to do,
it may do a call to its own \emph{sensing} method
(line~\ref{line:node:sense} ff.).  The sensor data, which for simplicity
is just a counter, is added to the \texttt{sendqueue} list together with
the sensor's id.  The \texttt{sendqueue} list contains all messages
waiting to be sent by the sensor.

If there are elements in the \texttt{sendqueue} list, the sensor's
\texttt{run} method may also call its own \texttt{sendOrForward} method
(line~\ref{line:node:send} ff.). A call to this method will result in the
broadcasting of the first message in the \texttt{sendqueue} list by a
call to the \texttt{broadcast} method of the network object. If both
sensing and sending is possible, then the node will perform a
non-deterministic choice between the two actions. If none of these
actions are possible, it will release the processor and wait.

When a sensor receives a message from another sensor, a call to the
\emph{receive} method is made by the network. If the sensor has not seen
this message before, it is added to the \texttt{received} list, and
additionally queued for re-sending.

\emph{Timed behavior} was added to the node class by modifying the
\texttt{sendOrForward} method.  When executing that method, the current
time is stored.  We specify that sending a message takes one unit of
time, so after broadcasting (and after synchronizing with the network to
ensure broadcasting has finished), line~\ref{line:node:timing} waits
until that amount of time has passed before terminating the method.  The
\texttt{sending} field serves as a mutual exclusion flag (only one
sending call can be active at any time); note how the cooperative
scheduling of Creol makes this style of programming both safe and easy
to understand.  Also note that blocking the sending method in this way
does not preclude the sensor from receiving messages, even while it is
waiting for the network to accept its message.

\begin{figure}
\begin{creol}
class SinkNode(network: Network)
implements Node
begin
  var noStored: Int := 0;
  var received: List[[Int,Int]] := nil;
  var lastReceived: Time;

  op init ==
    lastReceived := now

  op store(in data: [Int,Int]) == 
    noStored := noStored + 1

  with Network
    op receive(in data: [Int,Int]) ==
      if (lastReceived < now) then lastReceived := now end;
      if ~(data in received) then $\textsl{// store if not seen before}$
        received := received |- data;
        store(data;)
      end
end
\end{creol}
 \caption{The implementation of the sink node}
\label{fig:sink}
\end{figure}

The \texttt{SinkNode} class given in Figure~\ref{fig:sink} implements
the same interface as the sensor nodes, but has a different behavior.  The
major difference is that the sink has no \emph{run} method, and hence no
activity of its own.  The \texttt{store} method of the sink counts the
number of unique messages received, the \texttt{receive} method notes the time
when the last message was received.

\begin{figure}[t]
\begin{creol}
class BroadcastNetwork()
  implements Network
begin
  var nodesConns: Map[Node, List[Node]] := empty();
  var lastTransmission: Time;

  op init ==
    lastTransmission := now

  with Any
    op register(in node: Node, connections: List[Node]) ==
      nodesConns := insert(nodesConns, node, connections)

  with Node
    op broadcast(in data: [Int,Int]) == $\label{line:network:broadcast}$
      var rec: Node;
      var recs: List[Node] := nil;

      if caller in nodesConns then
        recs := get(nodesConns, caller)
      end;

      await now > lastTransmission; $\label{line:network:constraint}$
      lastTransmission := now;

      while ~isempty(recs) do
        rec := head(recs);
        recs := tail(recs);
        if rec /= caller then
          !rec.receive(data)
        end
      end
end
\end{creol}
\caption{The implementation of the network}
\label{fig:network}
\end{figure}

Figure~\ref{fig:network} shows the implementation of the network.  Its
behavior is implemented by the \texttt{broadcast} method
(line~\ref{line:network:broadcast} ff.).  We implemented different
behavior for that method in the case of multiple senders in a time slot:
\begin{itemize}
\item No message collision (broadcast all messages)
\item Collision sensing and re-send
\item Message loss
\end{itemize}
The \texttt{broadcast} implementation of Figure~\ref{fig:network} shows
the second behavior (re-sending messages).  Conceptually, re-sending
means that the sensor broadcasts the message until it is successful (no
collisions happened).  This is implemented by the field
\texttt{lastTransmission} in the network, which in conjunction with
line~\ref{line:network:constraint} only lets one \texttt{broadcast}
method call complete per time slot.  (We make the optimistic assumption
that the first broadcast actually succeeds -- i.e.\ the sensors'
antennas sense another transmission going on and the sensors abort their
own attempt at sending.)

The model's behavior for message collision can be adapted easily.  If
line~\ref{line:network:constraint} is removed, multiple sensors can send
at the same time. If line~\ref{line:network:constraint} is changed to
\begin{creol}
if lastTransmission = now then recs := nil end;
\end{creol}
all messages sent in a time slot after the first one are dropped.

\begin{table}[t]
  \centering
\begin{tabular}{|l|l|c|c|}
\hline
  \multirow{2}*{\textbf{Collision behavior}} 
  & \multirow{2}*{\textbf{Topology}} 
  & \textbf{No.\ received} 
  & \textbf{Timestamp of} \\ 
&&\textbf{by sink}&\textbf{last transmission}\\
\hline
\multirow{3}*{no interference} 
& linear & 12 & 14 \\
& mixed & 12 & 14 \\
& star & 12 & 2 \\
\hline
\multirow{3}*{resend} 
& linear & 12 & 60 \\
& mixed & 12 & 38 \\
& star & 12 & 12 \\
\hline
\multirow{3}*{drop} 
& linear & 0 & -- \\
& mixed & 2 & 4 \\
& star & 2 & 2 \\
\hline
\end{tabular}
  \caption{Timing results for different network behavior}
  \label{tab:results}
\end{table}

\begin{figure}[b]
  \centering
  \includegraphics[width=11cm]{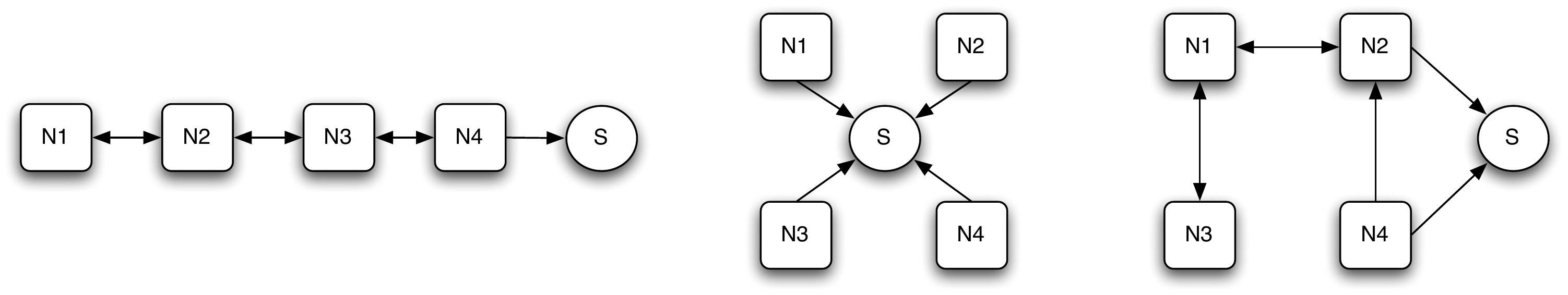}
  \caption{Network topologies: linear, star, mixed (from left to right)}
  \label{fig:topologies}
\end{figure}

Table~\ref{tab:results} shows the effect of these different network
behaviors on a network consisting of four sensor nodes sending three
messages each.  Three different topologies were simulated: the edge
cases of each node having a direct connection to the sink in a star
shaped network, and all nodes being arranged in a linear fashion, and a
more typical ``mixed'' network, with two nodes being able to reach the
sink, the other nodes having to send through them (see
Figure~\ref{fig:topologies}).  For the mixed topology, only two messages
manage to reach the sink at all in case of message loss upon collision;
interference without message loss more than doubles the time until all
messages reach the sink.  The linear configuration exhibits the worst
behavior, with potentially no message reaching the sink at all without
re-sending messages upon collision.

\section{Related Work}
\label{sec:related-work}

Actor systems have been used to model mobile ad-hoc networks, which are
similar to our biomedical sensor networks,
in~\cite{DBLP:conf/euc/DedeckerB04}, but this work does not include a
notion of time in its model.  Many approaches to timed modeling, for
example the UPPAAL tool~\cite{DBLP:conf/hybrid/BengtssonLLPY95} use
variants of timed automata~\cite{DBLP:journals/tcs/AlurD94}.  We decided
to take a different approach, namely to augment existing behavioral
models with timing properties.  Kyas and
Johnsen~\cite{DBLP:conf/fmco/KyasJ08} describe a timed extension to
Creol that is similar to the one proposed in this paper.  Their approach
is a bit more expressive, since time constraints can be expressed as
intervals in the model, compared to our approach, where only a minimum
passed time can be declared and the semantics do not enforce progress.
On the other hand, we give an operational semantics of timed Creol
instead of a denotational one, along with an interpreter that implements
``earliest possible completion'' semantics for the language.

\subsection{Real-time Maude}

Real-time Maude is an extension of Full Maude~\cite{Clavel02}
introducing a concept of time to support real-time rewrite
theories. It supports both discrete and dense time domains. New
datatypes for time are introduced, and the user must specify what
time model to use, or make a new one. In order to make time advance at
the same pace throughout the system, an encapsulation of the
configuration is required, and the sort \texttt{GlobalSystem} is
introduced as follows: 
\begin{maude}
op {_} : System -> GlobalSystem . 
\end{maude}
Time advances by applying a \emph{tick} rule.  A typical \emph{tick}
rule would be something like:
\begin{maude}
crl [tick] : {SYSTEM} => {delta(SYSTEM, R)} in time R if R <= mte(SYSTEM) .
\end{maude}
The \emph{delta} function propagates the time elapse throughout the
configuration and the \emph{mte} function calculates the maximum time
elapse. Both these functions must be defined by the user. \texttt{R} is
of type \emph{Time} and is a random number not greater than the maximum
time elapse of the system. If it is desired, the time may advance with a
fixed number of time steps by replacing \texttt{R} with a number, and
omitting the if part of the tick rule.

In contrast to Real-time Maude, our extension of Creol is targeting
distributed concurrent objects with a simple notion of abstract
discrete time.  This gives a lightweight formalization of time which
is strong enough to model time-outs as well as synchronized behavior
suitable for wireless communication.  In our setting the \emph{tick}
rule is simpler than in Real-time Maude in the sense that it can be
formulated without use of auxiliary functions depending on
user-defined equations. Thus the Creol programmer needs not understand the
technicalities of the tick rule nor the time model. 
The notion of time is fully predefined, and
the programmer's role is to use time-outs when modeling timed systems.

Our time model could in principal be implemented in Real-time Maude,
but the \emph{delta} and \emph{mte} equations would be complicated due
to the internal structure of Creol objects,
and take into account
time-outs, \emph{await} statements and blocking calls appearing in the
imperative code in the objects and also in the object's internal process
queue.

Previous work on wireless sensor modeling in Real-Time Maude includes
statistical model checking of wireless sensor network algorithms, 
and has demonstrated that  this approach can be used to 
detect flaws in non-trivial wireless sensor network algorithms
\cite{olveczky09journal,katelman08fmoods}.
The present work is not considering probabilistic modeling;
however, in current work towards probabilistic Creol semantics 
we are trying to exploit this approach.

\section{Conclusions}
\label{sec:conclusions}

We have presented an extension of \Creol with a lightweight model of
time which supports the modeling of radio communication and
time-dependent cooperative scheduling of tasks. Technically, this is
done by allowing read-only access to a global clock, and thereby time
related guards including lower and upper timer bounds. Currently, the
time model does not support continuous time.
Exploiting Creol's concept of asynchronous methods calls, time-outs
may be used to model the passing of time while blocking the processor,
as well as the passing of time while releasing the processor.  This
means that the passing of time can be tightly integrated into a \Creol
program. Due to the non-determinism in \Creol, it is easy to capture
distributed systems where the concurrent objects are progressing at
independent speeds.  This allows the modeling of ``best-effort''
systems where a time-out may be chosen some time after the given time
limit.

The model presented here is a simplification of earlier timed
versions of Creol~\cite{johnsen07fmco}, where both local and global time
were modeled. Local time allowed a more direct implementation of objects
progressing at independent speeds and of timing of communication events
\cite{johnsen07fmco}.  The present model is without local clocks and
gives the programmer more freedom to model the progression of time.  Another
advantage is the ease with which timing information can be added to
existing models.  The model in Section~\ref{sec:example} started out
as an untimed model; augmenting it with timing constraints was a matter
of adding less than 10 lines of code in total.  We believe that this ease
of expressing timing constraints is a valuable tool for the modeler.

Our semantic model of timed Creol
is not suitable for modeling of
systems where clock drift is relevant. It  is suitable when one
may assume perfect clock synchronization between nodes.
Experiences with the more low-level semantics with local clocks
given in \cite{johnsen07fmco} 
showed that this model was difficult to analyze
with the Maude tools
due to the large state space.
The motivation behind the current approach was to 
achieve a model with a smaller state space,
and still expressive enough to cover interesting Creol models.
The investigation of the applicability of Maude analysis tools  is still future work.

The simplicity of our model seems promising with respect to automated
analysis including state space exploration and model checking
tools. In fact, timed simulation of an untimed model will in many
cases have less states than the corresponding untimed simulation, due
to the constraints imposed by the global clock.
A main motivation behind the design of Creol is simplicity of
reasoning and simplicity of composition rules. In contrast to the
standard multi-threaded concurrency model of object-oriented systems,
\Creol has a compositional proof system \cite{deboer07esop}.  For
partial correctness, the simplicity of the reasoning system for \Creol
can be preserved for the extension of \Creol with time-outs.  The
time-outs can be handled in the reasoning system as a special kind of
\emph{await} statements, using the approach outlined in
\cite{dovland05swste}.

The extension of \Creol with time does not depend on very specific
features of the language.  \Creol was chosen as a base language to
allow exploitation of its executable Maude interpreter by extending it
with time and experimenting with case studies.  The presented
lightweight model of time seems applicable to the wider setting of
distributed concurrent object or actor-based systems where
communication is by message passing.

\paragraph{Acknowledgments}
The original, untimed BSN sensor network model in Creol was created by
Wolfgang Leister and Xuedong Liang.  The authors wish to thank S.\
Lizeth Tapia Tarifa and the anonymous reviewers for valuable comments.

\bibliographystyle{eptcs}

\end{document}